\documentclass[a4paper,11pt]{article}
\pdfoutput=1

\usepackage{jcappub}
\usepackage[english]{babel}
\usepackage{amsmath,amssymb}
\usepackage{upgreek}
\usepackage{lineno}
\usepackage{xspace}
\usepackage{graphicx}
\usepackage{booktabs}
\usepackage[capitalize]{cleveref}
\usepackage{siunitx}
\usepackage{xcolor}

\graphicspath{{figures/}} 

\newcommand{\xmax}[0]{\ensuremath{X_{\rm max}}}
\newcommand{\rt}[0]{\ensuremath{t_{1\mspace{-2mu}/\mspace{-1mu}2}}}

\title{\boldmath Search for photons above $\text{10}^\text{19}$\,eV with the surface detector of the Pierre Auger Observatory}
\date{\today}


\arxivnumber{number not yet known}

\author[71]{P.~Abreu,}
\author[53,51]{M.~Aglietta,}
\author[1]{I.~Allekotte,}
\author[69]{K.~Almeida Cheminant,}
\author[8,12]{A.~Almela,}
\author[78]{J.~Alvarez-Mu\~niz,}
\author[78]{J.~Ammerman Yebra,}
\author[53,51]{G.A.~Anastasi,}
\author[85]{L.~Anchordoqui,}
\author[8]{B.~Andrada,}
% ...

\affiliation[1]{Centro At\'omico Bariloche and Instituto Balseiro (CNEA-UNCuyo-CONICET), San Carlos de Bariloche, Argentina}
% ...
\affiliation[8]{Instituto de Tecnolog\'\i{}as en Detecci\'on y Astropart\'\i{}culas (CNEA, CONICET, UNSAM), Buenos Aires, Argentina}
% ...
\affiliation[12]{Universidad Tecnol\'ogica Nacional -- Facultad Regional Buenos Aires, Buenos Aires, Argentina}
% ...
\affiliation[51]{INFN, Sezione di Torino, Torino, Italy}
% ...
\affiliation[53]{Osservatorio Astrofisico di Torino (INAF), Torino, Italy}
% ...
\affiliation[69]{Institute of Nuclear Physics PAN, Krakow, Poland}
% ...
\affiliation[71]{Laborat\'orio de Instrumenta\c{c}\~ao e F\'\i{}sica Experimental de Part\'\i{}culas -- LIP and Instituto Superior T\'ecnico -- IST, Universidade de Lisboa -- UL, Lisboa, Portugal}
% ...
\affiliation[78]{Instituto Galego de F\'\i{}sica de Altas Enerx\'\i{}as (IGFAE), Universidade de Santiago de Compostela, Santiago de Compostela, Spain}
% ...
\affiliation[85]{Department of Physics and Astronomy, Lehman College, City University of New York, Bronx, NY, USA}
% ...
\affiliation[c]{Louisiana State University, Baton Rouge, LA, USA}
% ...

\emailAdd{spokespersons@auger.org}

\abstract{We use the surface detector of the Pierre Auger Observatory to search for air showers initiated by photons with an energy above $10^{19}$\,eV.  Photons in the zenith angle range from 30$^\circ$ to 60$^\circ$ can be identified in the overwhelming background of showers initiated by charged cosmic rays through the broader time structure of the signals induced in the water-Cherenkov detectors of the array and the steeper lateral distribution of shower particles reaching ground. Applying the search method to data collected between January 2004 and June 2020, upper limits at 95\% CL are set to an $E^{-2}$ diffuse flux of ultra-high energy photons above $10^{19}$\,eV, $2{\times}10^{19}$\,eV and $4{\times}10^{19}$\,eV amounting to $2.11{\times}10^{-3}$, $3.12{\times}10^{-4}$  and $1.72{\times}10^{-4}$~km$^{-2}$~sr$^{-1}$~yr$^{-1}$, respectively. While the sensitivity of the present search around $2 \times 10^{19}$~eV approaches expectations of cosmogenic photon fluxes in the case of a pure-proton composition, it is one order of magnitude above those from more realistic mixed-composition models. The inferred limits have also implications for the search of super-heavy dark matter that are discussed and illustrated.}

\begin{document}
\modulolinenumbers[5]
\setlength\linenumbersep{0.8ex}
\maketitle
\flushbottom

\section{Introduction}
\label{sec:intro}

Photons with energies above $10^{19}$\,eV can be produced by $\pi^0$ decays subsequent to the interactions of ultra-high energy cosmic rays (UHECRs) with the photon fields or the dust permeating the source environments, or the background photon fields in the extragalactic space. The resulting photon fluxes are attenuated over distances of ${\sim}$10\,Mpc by $e^\pm$ pair productions subsequent to the interactions of these photons with those of the cosmic-background, see e.g.~\cite{Coppi:1996ze,Heitler:2018,Nikishov:JETP1962}. Consequently, the detectable volume of photon sources encompasses only the local Universe while the UHECR interactions within this volume produce a guaranteed diffuse photon flux. 

The cosmogenic photon flux depends on the nature of the UHECRs. The hadrons that cause the creation of the $\pi^0$ mesons, through the process of resonant photopion reaction, must have energies typically ten times higher than the secondary photons. Such hadrons can be primary proton CRs, or secondary ones produced from the photo-disintegration of nuclei interacting inelastically with a cosmic-background photon, which leads to the production of nucleons inheriting the energy of the fragmented nucleus divided by its atomic number. Given the steepening of the UHECR intensity with energy, photons are thus more efficiently produced  above \SI{e19}{eV} by UHE protons. Several mass-sensitive observables are however providing evidence that the mass composition of UHECRs is gradually getting heavier above $10^{18.3}$\,eV and is, in particular, not compatible with a pure-proton composition~\cite{LongXmax2014, Auger_Mass2_2014,Auger_Mix_2016,Auger_Delta2017,TA:PhysRevD.99.022002,Abbasi_2018,AugerTA_MassCompJoint2019}. This is in line with the absence of copious fluxes of cosmogenic photons and neutrinos with energies ranging from GeV to EeV, as reported in~\cite{Fermi_EGR2010} from the extragalactic gamma-ray flux at GeV energies and in~\cite{IceCube_CosmogNu2016,Auger_NuOrigin2019} from neutrino searches above a hundred of PeV. These results provide important constrains on the 
sources of protons at UHE~\cite{Berezinsky_CascadePhotons2016,Heinze_NuDip2016,Supa_PhotonProton2016,AlvesBatista_NuPhoton2019, Muzio_UFA2019}. Even though scenarios based on a mixed composition of UHECRs are more demanding in terms of exposure to photons, the search for these emblematic messengers is thus complementary to complete the multi-messenger approach aimed at understanding the non-thermal processes producing UHECRs in the Universe. 

While the search for photons above \SI{e19}{eV} is of primary importance to decipher further the origin of UHECRs, the detection of photons of even higher energies, above \SI{e20}{eV}, would open an unexpected window, revealing either new physics such as Lorentz invariance violation~\cite{COLEMAN1997249,BHATTACHARJEE2000109,BIETENHOLZ2011145,PhysRevD.86.103006,PhysRevD.89.123011,0034-4885-77-6-062901,1475-7516-2017-05-049} or signatures of axion mixing models~\cite{1475-7516-2010-04-013,Fairbairn2011_Axions}, or some new particle acceleration never seen or imagined until now. At the same time, the detection of a flux of UHE photons could be compelling evidence for dark matter (DM) composed of super-heavy particles. In cosmological models with an inflation phase, such particles never at thermal equilibrium could have been produced at reheating after inflation through mechanisms involving gravitation~\cite{Berezinsky:1997hy,Chung:1998zb,Garny:2015sjg}. Despite being metastable particles, they can decay through non-perturbative effects into standard-model particles~\cite{Kuzmin:1997jua,Berezinsky:1997hy}, and hence produce detectable secondaries such as nucleons and photons. Of particular interest would thus be the detection of UHE photons from regions of denser DM density such as the center of our Galaxy. The limits on such photon fluxes translates into constrains on the lifetimes and masses of DM particles~\cite{Aloisio_SHDM2015,Alcantara_SHDM2019,Kachelriess_NuSHDM2018,Ishiwata_SHDM2020}, or even on the particle-physics properties of the dark sector~\cite{PierreAugerCollaboration:2022tlw,PierreAugerCollaboration:2022wir}.

In this work we update the search for UHE photons above $10^{19}$~eV using the surface detector (SD) array of the Pierre Auger Observatory~\cite{Auger_NIM_2015}. Compared to previous analyses~\cite{ABRAHAM2007155,ABRAHAM2008243,BleveICRC2015,AugerPhoton_icrc2019}, this work benefits from the increased exposure cumulated from January 2004 to June 2020, as well as from a refined search method and data selection. The search for photons presented here complements and extends previous searches using data from the Pierre Auger Observatory at lower energies~\cite{Savina:2021zrn,HECOPhotonPaper}.
The paper is organized as follows.
In~\cref{sec:EAS-PAO}, general features of photon induced extensive air showers are presented focusing on the differences expected with respect to the bulk of showers initiated by nuclei that constitute the background for the search. The Pierre Auger Observatory is also briefly described, with a more specific emphasis on the SD array used in this analysis. 
The discriminating variables aiming at identifying photon showers are introduced  in~\cref{sec:discrim_variables}. The Monte Carlo simulations of photons used in the analysis, the photon energy scale, and the analysis method combining the discriminating variables to extract photons from the bulk of events are detailed in~\cref{sec:Analysis}. The results of the photon search and the upper limits to the diffuse flux of UHE photons are presented in~\cref{sec:results}. Finally, some astrophysical implications of the results are discussed in~\cref{sec:discussion}.


\section{Photon showers at the Pierre Auger Observatory}
\label{sec:EAS-PAO}

The Pierre Auger Observatory is a ground-based instrument designed to detect the extensive air showers (EAS) induced in the atmosphere by UHECRs. 
We briefly discuss here the main features of the photon-induced showers compared to nucleus-induced ones and how the showers are detected and reconstructed at the Pierre Auger Observatory. A more detailed description can be found elsewhere for extensive air showers~\cite{EAS_models_review,RissePhoton_2007}, the Observatory~\cite{Auger_NIM_2015} and the reconstruction of EAS~\cite{SDreco_2020}.

 \subsection{Main features of photon showers}
\label{subsec:Photons}
 \begin{figure}[t] 
	\centering
  \includegraphics[width=0.5\textwidth]{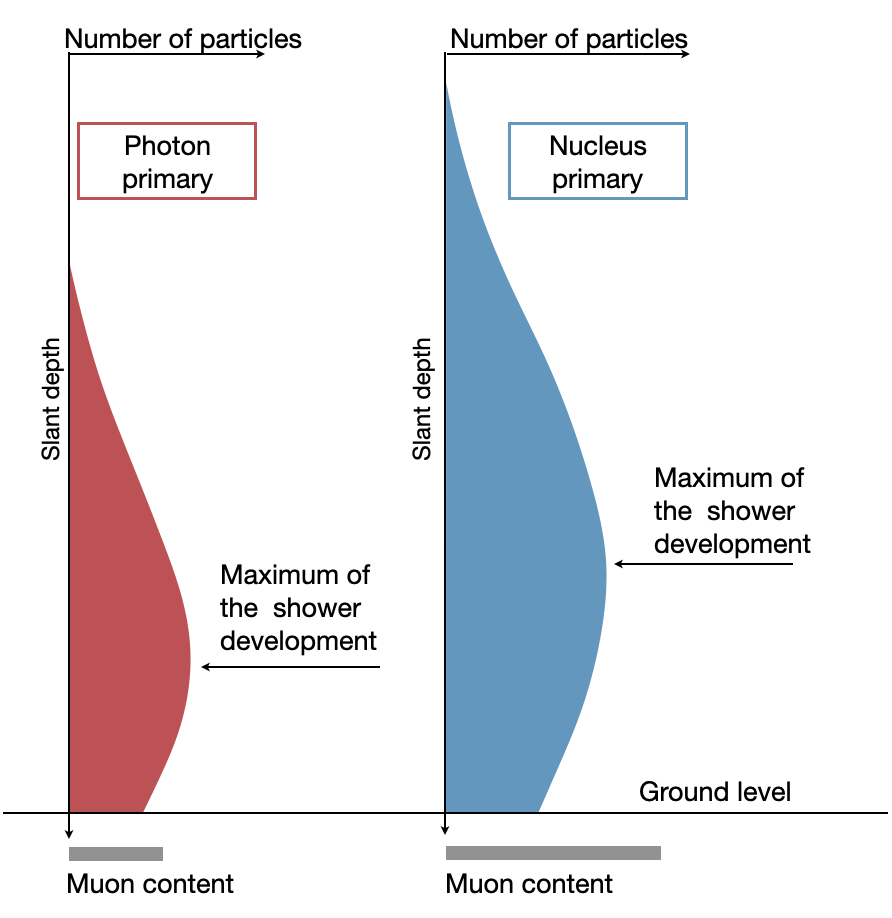}
	\caption{Main features of photon- and nucleus-induced showers.}
	\label{fig:EASschematic}
\end{figure}

Air showers initiated by high-energy photons in the atmosphere differ significantly from those from nuclei.
For a photon-induced shower, the first interactions and generations are purely electromagnetic, since the radiation length is more than two orders of magnitude smaller than the mean free path for photo-nuclear interactions. Yet, the development of the shower is delayed by the typically small multiplicity of electromagnetic interactions. Thus the maximum development of the shower is reached at a slant atmospheric depth $X_{\rm max}$ larger for photon primaries than for nuclei, with a difference of $\simeq 200 \ \rm{g\ cm^{-2}}$ between photons and protons at $10^{19}$~eV and even larger between photons and heavy nuclei.

Processes in the cascade give rise in general to secondary particles inheriting from a moderate transverse momentum. Most of the high-energy particles are thus collimated along the shower axis. However, low-energy particles can extend in a halo up to a few kilometers from this core. In particular, the electromagnetic part of the halo increases with the slant depth $X$ before decreasing when the core is no longer active for regenerating the cascade. Overall, the steepness of the lateral distribution decreases with  $X$ so as to get flatter through the shower development, and the fall-off with the distance to the axis of the shower depends on the primary mass of the cosmic rays. At ground level, the steepness is thus relevant to distinguish between  nucleus-induced showers and photon-induced ones.

Since the mean free path for photo-nuclear interactions is much larger than the radiation length, the transfer of energy to the hadron and muon channels is reduced hence only a small fraction of the electromagnetic component in a photon-induced shower is injected into the hadronic cascade. Showers induced by photons are thus characterized by a lower content of muons: on average, simulations show that photon showers have nearly one order of magnitude less muons than  proton showers of the same energy. 

These main features of photon showers, depicted in \cref{fig:EASschematic}, are amplified by the Landau-Pomeranchuk-Migdal (LPM) effect~\cite{Landau:1953um,Migdal:1956tc} resulting in a suppression of the bremsstrahlung and pair-production cross sections.

The picture of UHE photon showers is supplemented by accounting for the influence of the magnetic field of the Earth, which can allow for the conversion
of photons into an $e^\pm$ pair before they enter the upper atmosphere (``preshowering'' effect \cite{McBreen_preshower_PRD1981}).
The resulting showers are a superposition of cascades initiated by lower energy electrons and photons, giving rise to smaller $X_{\rm max}$  values on average than photon showers of the same energy not affected by the preshowering, and, as a consequence, a reduced separation in the average $X_{\rm max}$ from nucleus-induced showers.

\subsection{The Pierre Auger Observatory and the events collected with the surface detector}
\label{subsec:PAO}

The Pierre Auger Observatory is located in the province of Mendoza, Argentina, at 1400\,m a.s.l. -- corresponding to an atmospheric overburden of $\simeq$ \SI{875}{g/cm^2}. It is designed as a hybrid cosmic-ray detector using two proven techniques to measure the properties of EAS by observing their longitudinal development in the atmosphere with a fluorescence detector (FD) and their lateral spread at ground level with a surface detector (SD) array. The search for photons presented here makes use of data collected only with the SD, which operates with $\simeq$100\% duty cycle.

The SD array consists of a triangular grid of about 1600 water-Cherenkov detectors (WCDs), with a spacing of 1500~m, covering a total area of 3000\,km$^2$. Each WCD is a cylinder with a surface area of 10~m$^2$ and a height of 1.2~m, holding 12 tonnes of ultrapure water viewed by three $9''$ photomultipliers (PMTs). These detect the Cherenkov light emitted in water by charged particles and the $e^\pm$ pairs produced in water by secondary photons reaching the ground. The signals from the PMTs are digitized using 40 MHz 10-bit flash analog-to-digital converters (FADCs). 
The signals are normalized to the signal obtained for a vertical muon and expressed in vertical equivalent muons (VEM).
Data are collected in real time by searching for temporal and spatial coincidences at a minimum of three WCD locations to build the event triggers. When this occurs, the pulse amplitude and time of detection of signals are obtained from the FADC data of the PMTs. The data quality is checked by both an on-line and a long-term continuous monitoring of the detectors. 

The arrival direction of the primary particle initiating the EAS is reconstructed  using  the start time of the signals recorded in individual detectors and is determined with a resolution of $\simeq 1^\circ$.
To estimate the energy of the primary particle, the total signal amplitude, integrated in time, of each triggered detector, $S_i$, is used.

The lateral extension of the showers, also known as the lateral distribution function (LDF), is measured at the ground level. The signal deposit $S_i$ in each WCD is adjusted by scaling the normalisation of an average LDF in a fitting procedure so as to best reproduce the observed signal amplitudes~\cite{SDreco_2020}.
The energy estimator is then the signal interpolated at 1000~m from the shower axis, $S(1000)$.
The attenuation of $S(1000)$ in the atmosphere for showers with same energy but different zenith angle is accounted for with the constant intensity cut method~\cite{CIC}. The final value of $S(1000)$ is calibrated with the quasi-calorimetric measurement of the primary energy provided by the fluorescence detector for a subset of hybrid events~\cite{Aab:2020gxe}.

According to the procedure described, the energy  $E_{\rm had}$ assigned to each event is a function $f_{\rm had}( S(1000),\theta )$, whose parameters are calibrated with data and represents the correct energy scale for nuclear primaries.
Since  photons are characterised by an $X_{\rm max}$ and a muonic content significantly different from the bulk of data, $E_{\rm had}$ is not suitable to provide an accurate estimate  of the energy of photon showers, resulting indeed in an overestimate of more than a factor two. An alternative procedure is therefore required to assign the correct photon energy scale, as explained in~\cref{subsec:photon_energy}.

\section{Observables from the Surface Detector}
\label{sec:discrim_variables}

\begin{figure}[t] 
	\centering
  \includegraphics[width=0.6\textwidth]{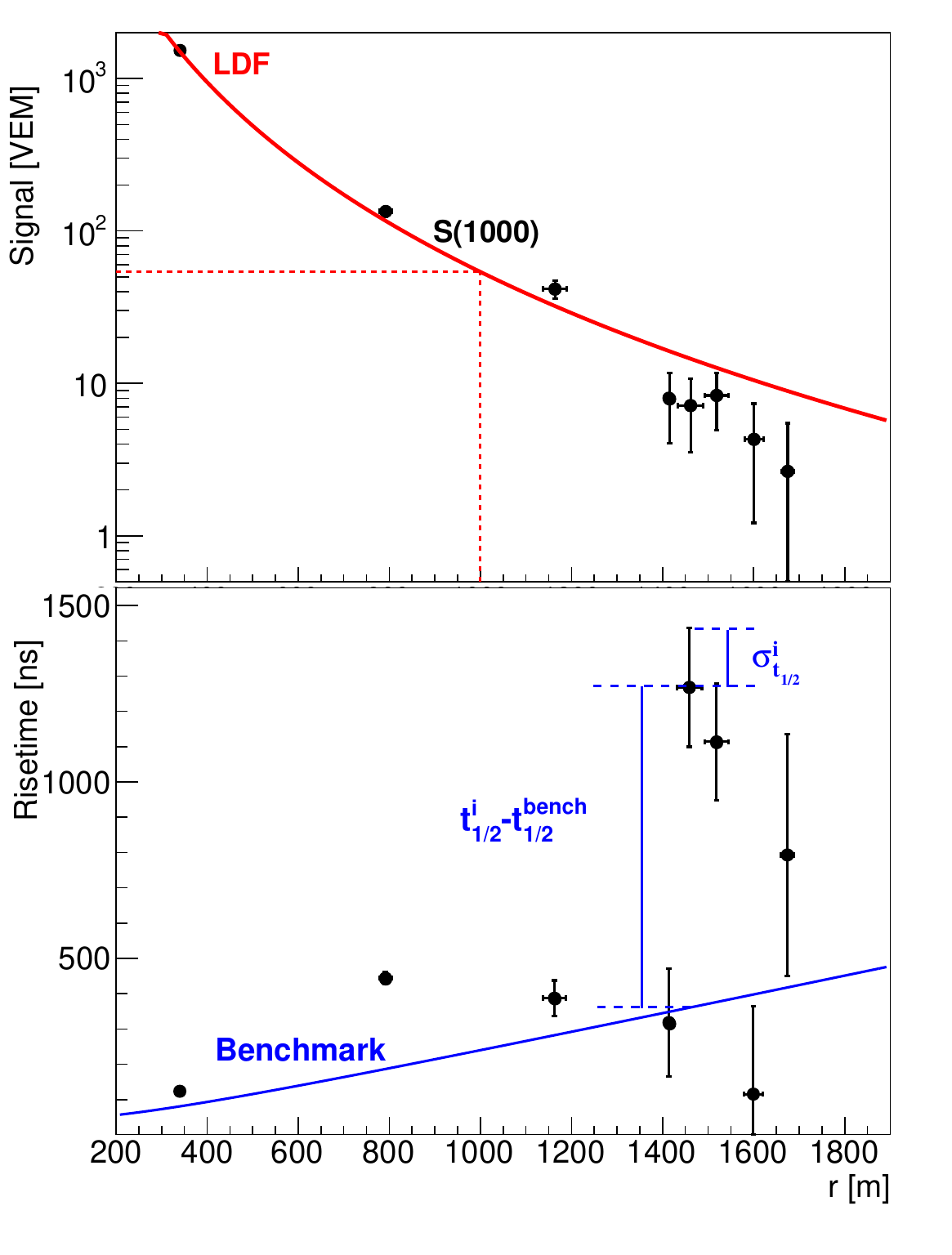}
	\caption{A shower induced by a simulated photon of 39 EeV with a zenith angle of 44$^\circ$: 
(top)~the lateral distribution of signals in the WCDs is steeper than the LDF obtained from data (solid
line) and (bottom) the risetime $\rt^{i}$ of the signal in the WCDs is larger than the average data benchmark $\rt^{\rm bench}$ (solid line). The vertical bars in the bottom panel represent the sampling fluctuations  $\sigma^{i}_{\rt}$ of the risetime, parameterized for data. They are drawn separately for each station as the parameterization is a function of the total signal in the WCD and cannot be represented as a single band around the risetime benchmark.}
	\label{fig:discrim_param}
\end{figure}

The main characteristics of an air shower for the identification of the nature of the primary particle are its \xmax{} value and the muonic content (\cref{subsec:Photons}).
While  \xmax{} can be measured directly with the FD, this is not possible with the SD that detects the secondary particles of the air shower reaching the ground. The muon content, as well, cannot be estimated 
with WCDs, which record an aggregate signal from muons and other electromagnetic shower components.
We use instead two robust data-driven variables describing the spread in time of the shower front and the steepness of the lateral distribution of time-integrated signals, sensitive both to the depth of the shower maximum and the muonic content of the EAS.
The strategy to search for photons is to identify, in the bulk of events detected, showers that depart significantly from the average behaviour of data in the direction expected for photon primaries, i.e. events with a larger spread in the arrival times of the secondaries and a steeper LDF.

The use only two variables for the classification is aimed at designing the photon candidate selection using a small sample of data instead of a large sample of simulated hadron-induced showers (\cref{sec:Analysis}).
A study of several mass dependent observables led us to the choice of the two variables described in the following on the base of their robustness and best classification performance.

\subsection{Signal risetime}

The spread in arrival times of secondary particles in individual WCDs can be measured through the \textit{risetime}  defined as the time at which the integrated signal in the FADC time trace rises from 10\% to 50\% of its total value. The risetime is increased by a larger contribution of the electromagnetic component as secondary photons and electrons undergo scattering and attenuation in the atmosphere, unlike muons, which are concentrated in time close to the shower front. It also increases when the difference in depth between \xmax{} and the observation level becomes smaller, for geometrical reasons~\cite{Aab:2017cgk}.
Being sensitive to both the deeper \xmax{} and the limited muon content of photon showers, the risetime is a suitable variable for the search of photons.

A ``Data Benchmark'' is produced to describe the average risetime of data as a function of the zenith angle and of the distance to the shower axis, following a procedure similar to the one described in detail in \cite{Aab:2017cgk}.
A correction for asymmetries in the observed risetime is obtained from data, accounting for the fact that for non-vertical air showers, the ground level observation corresponds to different stages of development (or ``age'') of the shower for geometrical reasons. In the following, we will denote the asymmetry corrected risetime as \rt.
Sampling fluctuations, $\sigma_{\rt}$, are also estimated from the data, using the difference between the measurements of SD doublets (a WCD in the regular SD grid plus a second one off-grid deployed close to it) or SD pairs (two WCDs in the same event with similar distance from the axis and total signal).

The risetime benchmark $\rt^{\rm bench}$ and  $\sigma_{\rt{}}$ describe the average thickness of the local shower disk for the bulk of cosmic rays detected with the SD. For each triggering detector in an event, the quantity
\begin{equation}
\delta_{i} = \frac{(\rt^{i} - \rt^{\rm bench})}{\sigma^{i}_{\rt} }
\label{eq:deltai}
\end{equation}
is then providing a measurement of the deviation of the risetime in the $i$-th WCD from the data benchmark in units of expected standard deviation.
An example of $\delta_{i}$ values for one simulated  photon shower event is shown in~\cref{fig:discrim_param}, where large departures from the benchmark curve are observed.

We can therefore define an observable that measures the departure of an individual event from the data-averaged behaviour of air-showers as 
\begin{equation}
\Delta = \frac{1}{N}\sum_{i=1}^N {\delta_{i}},
\label{eq:delta}
\end{equation}
where $N$ is the number of the triggered detectors in the event.
$\Delta$ is expected to average to zero for data by construction
and to be  positive for air showers initiated by photons.

\subsection{Steepness of the lateral distribution of signals}

The reduced muon content of photon showers with respect to data
produces, as already mentioned, a steeper LDF of the signals in the detectors at ground level.
At large distances from the axis, photon showers thus produce typically smaller signals than expected from the data LDF. This is illustrated in the top panel of~\cref{fig:discrim_param} where the LDF function is
\begin{equation}
f_{\mathrm{LDF}}(r)= S(1000) \ \left(\frac{r}{r_{\mathrm{opt}}}\right)^\beta  \ \left(\frac{r+r_s}{r_{\mathrm{opt}}+r_s}\right)^{\gamma+\beta},
\end{equation}
with $r_{\rm opt}= 1000$~m, $r_s = 700$~m and $\gamma$ and $\beta$ are parameterized as a function of $S(1000)$ and $\theta$ to describe the average behaviour of data~\cite{SDreco_2020}. 

We define an observable $L_{\rm LDF}$ measuring the departure of the observed signals from the average data LDF as the logarithm of the average ratio between the SD signals and $f_{\rm LDF}(r)$:
\begin{equation}
L_{\rm LDF} = \log_{10}\left( \frac{1}{N} \sum_{i=1}^N \frac{S_i}{f_{\rm LDF}(r_i)}\right),
\end{equation}
where $S_i$ is the total signal of the $i$-th detector and $i$ runs over the $N$ detectors with radial distance from the shower axis $r_i > $1000\,m, where the signal for photon showers is expected to be lower than $f_{\rm LDF}$. 
$L_{\rm LDF}$ is expected to be close to 0 for data, as by construction the LDF function is built to describe the data average behaviour, and negative for photons.

\section{Analysis method}
\label{sec:Analysis}
The data collected with the SD of the Pierre Auger Observatory between January $1^\text{st}$ 2004 and June $30^\text{th}$ 2020  are used for the analysis described here. The expected physical differences between signal and background  are quantified using simulations of air showers initiated by photons. The background of showers induced by nuclei is not simulated; instead a fraction of the data set is used as a burn sample to define the selection for candidate events. The search sample will consist of the remaining events.

\subsection{Monte Carlo simulations of photon showers}
\label{subsec:Library}

A set of 27,000 photon showers is simulated using CORSIKA~\cite{Heck:CORSIKA} with EPOS-LHC~\cite{PhysRevC.92.034906} as the high-energy generator of hadronic interactions and FLUKA~\cite{FLUKAWEB,FLUKA1,FLUKA2} as the low-energy interaction model. 
Since showers initiated by photons are almost purely electromagnetic in nature, no dependence is expected on the choice of the hadronic interaction model \cite{ABRAHAM2008243}.

The energy distribution of simulated showers follows a $\Phi_{\rm gen }(E) \propto E^{-1}$ law  in the  energy range $10^{18.5} - 10^{20.5}\,$eV; the arrival directions are distributed in zenith angle $\theta$ between  $20^\circ$ and $70^\circ$ according to a $\cos^2(\theta)$ distribution to simulate an isotropic flux impinging on a flat surface at ground. Any arbitrary energy spectrum other than the one simulated can be reproduced by weighting the simulated events with $w(E) = \Phi(E)/\Phi_{\rm gen}(E)$.

Both the  LPM effect and geomagnetic cascading (preshowering), described in~\cref{subsec:Photons}, are considered in the simulations \cite{Homola_preshower_APP2007,Homola:preshower2013}.

To reduce the computational resources needed for the simulations, a thinning algorithm is used \cite{KOBAL2001_thinning} with a thinning factor $t_{f} =10^{-6}$. The resulting distributions of particles at ground are de-thinned at the stage of the simulation of the detector response according to the statistical method described in \cite{BILLOIR2008_unthinning}. 

The  response of the SD array is simulated using the \mbox{$\overline{\textrm%
{Off}}$\hspace{.05em}\protect\raisebox{.4ex}%
{$\protect\underline{\textrm{line}}$}}\xspace
package 
\cite{Argiro:2007qg} providing as output simulated events in the same format as real events. Each CORSIKA shower is used five times, placing randomly the shower footprint on the area of the SD array.

\subsection{Photon energy scale}
\label{subsec:photon_energy}

To account for the different energy scale of photon showers (see~\cref{subsec:PAO}), we retain the standard reconstruction of the shower direction and energy estimator $S(1000)$ and replace $E_{\rm had}$ with a new function of $S(1000)$ and $\theta$ calibrated with photon simulations.

\begin{figure}[t]
	\centering
    \includegraphics[width=0.49\textwidth]{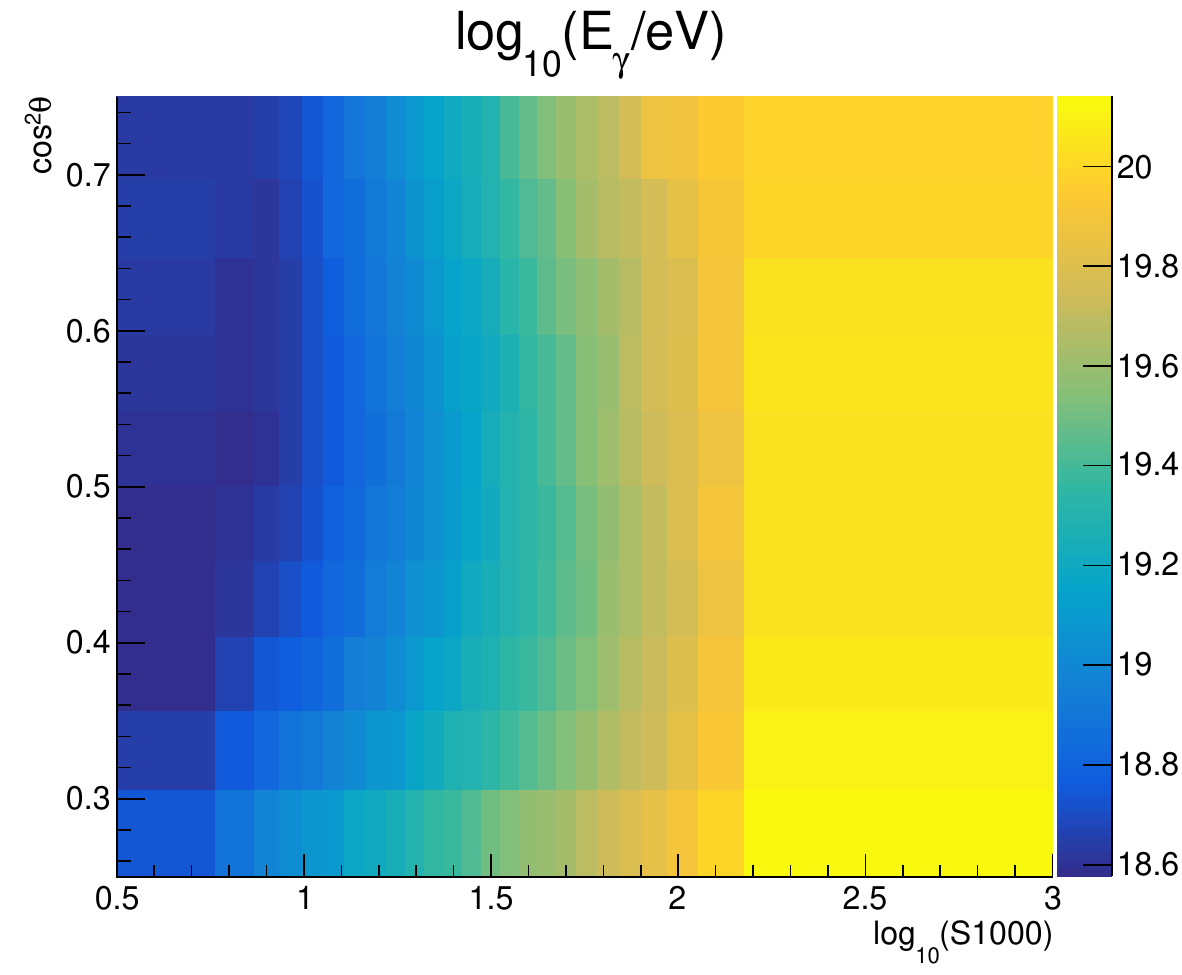}
    \includegraphics[width=0.49\textwidth]{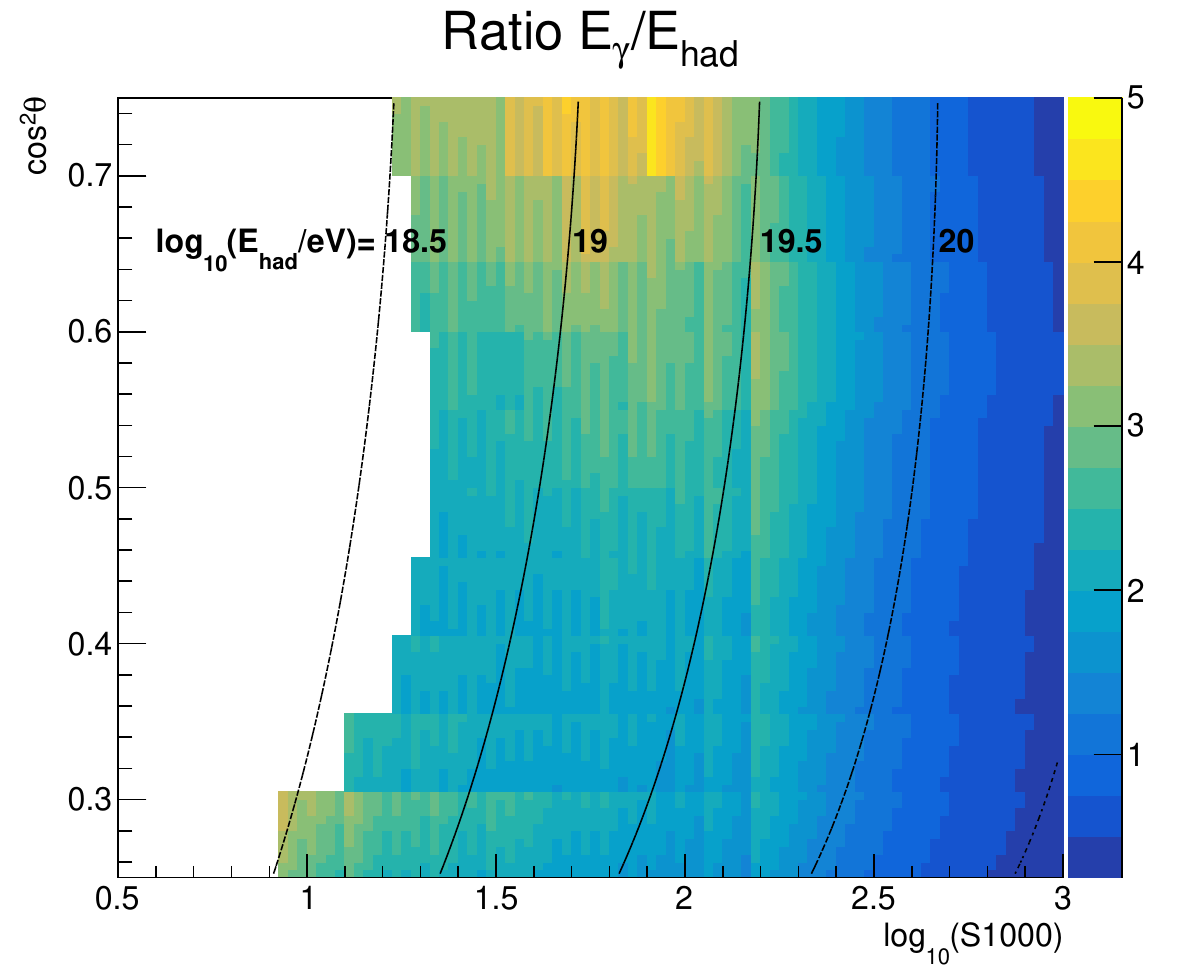}
	\caption{Look-up table used to assign the photon energy $E_{\gamma}$ (left) and ratio of $E_\gamma$ to the energy $E_{\rm had}$ calibrated with data and representing the correct scale for nuclear primaries (right). The ratio is shown only for $E_\gamma > 10$~EeV, corresponding to the range studied in this work. The lines in black show the contours corresponding to four values of $E_{\rm had}$ to facilitate the comparison of the energy scales.}
	\label{fig:ephoton}
\end{figure}

Taking advantage of the large statistics of simulated events, a look-up table is built with the mean value of the logarithm of the Monte Carlo true  energy of photons, $L_E={\log_{10}(E/{\rm eV}})$, in bins of $S(1000)$ and $\theta$. Only non-preshowering simulated events triggering the SD array are used and the photon simulations are weighted to a reference spectrum $\propto E^{-2}$. The table, shown in \cref{fig:ephoton},  serves as the desired function of $S(1000)$ and $\theta$: for each event the assigned photon energy $E_\gamma$ is $10^{L_E}$~eV, where $L_E$ is the tabulated value for the bin to which the event belongs.

\subsection{Event selection}
\label{subsec:cuts}

To ensure events with a good reconstruction and optimize the photon search, we exclude lightning events and we select only events in which:
\begin{itemize}
\item the detector with the highest signal is surrounded by a hexagon of six stations that are fully operational,
\item the reconstructed zenith is in the range $30^\circ-60^\circ$,  
\item the reconstructed energy in hadronic scale is  $E_{\rm had} >10^{18.5}$~eV.
\end{itemize}
The second condition ensures that the majority of selected photon-induced  showers reach their maximum development before being detected, as $\langle X_{\rm max} \rangle$ exceeds the vertical atmospheric depth at the Observatory site already at \mbox{$E_\gamma > 10^{19}$~eV} and increases with increasing $E_\gamma$.
The latter condition reduces the background from low energy showers initiated by nuclei while not impacting the selection efficiency for photons above 10~EeV when combined with the observable-related cuts.

The timing and shape of the signal acquired by each PMT of the WCDs is of paramount importance for the  risetime observable $\Delta$ defined in \cref{eq:delta}.  
The quality criteria applied to the SD events include the filter of PMTs with hardware and electronics problems, identified by the continuous monitoring of the SD.
Minor and very short-lived problems in the PMT operation can fall within the tolerance of the monitoring filter as it is designed to avoid effects on the standard reconstruction. In the case of the photon searches, however, it is also important to consider the rare cases of events that are already in the highest end of the $\Delta$ distribution being wrongly assigned a larger value because of a single PMT not working properly: even if they do not impact the mean and standard deviation of the $\Delta$ distribution, they can potentially be misclassified as photon candidates.
Three additional, stricter, filters have been therefore designed for the photon searches to exclude from the calculation of $\Delta$ individual PMTs  with minor and short-lived malfunctions of the same type addressed by the standard monitoring filter: an oscillating baseline pattern, significant afterpulses, or a non-physically slow decrease of their signal due to sudden fluctuations in the VEM peak or Dynode/Anode ratio. For the first problem the tolerance threshold for the baseline fluctuations is slightly lowered with respect to the standard monitoring, while to search for the other two effects each PMT trace is compared with the traces of the other two PMTs in the same WCD, on a event-by-event base, to identify anomalies not compatible with the normal operation of the PMT in the form of excess signal late in the trace or slow signal decrease. In this cases only the individual PMT is removed and the risetime for the WCD trace is recomputed using the other PMTs, with a negligible effect on the number of events selected,  amounting to 24 events less than without additional filters. Only 11 events have their $\Delta$ value significantly reduced (i.e. with a change $>$1 in absolute value, where the units are standard deviations of the sampling fluctuations of the risetime by definition of $\Delta$)\footnote{We verified a posteriori, after having performed the photon search, that 2 of the 11 events would have been indeed wrongly classified as photon candidates if the additional PMT filters were not in use. }.

With the selection criteria described above, the set of air showers detected with the SD in the time period of interest consists of 105,064 events.

To guarantee a good quality of the observables defined in~\cref{sec:discrim_variables}, additional criteria are applied:
\begin{itemize}
\item only WCDs with non-saturated signal $>6$~VEM and radial distance in the range $600-2000$~m are considered in the calculation of $\Delta$, the event being selected when the resulting number of selected stations $N \geq 4$,
\item  a minimum of one WCD above radial distance \SI{1000}{m} is required to compute $L_{\rm LDF}$.
\end{itemize}
The search is restricted to $E_{\gamma} >10^{19}$~eV, corresponding to the energy at which photons are detected by the SD with a trigger efficiency close to 100\%. 

These criteria reduce the set of selected data to 48,947 events.

The resolution on $E_\gamma$ for simulated photon events fulfilling the same selection criteria is almost constant over the energy range considered, being $\simeq 30\%$ with no bias for non-preshowering photons from a $E^{-2}$ spectrum. Showers undergoing preshowering in the geomagnetic field, on the other hand, are characterized by a shallower depth of the shower maximum with respect to non-preshowering ones of the same primary energy. This results in an underestimate of their energy of 30\% using the look-up table constructed as described in \cref{subsec:photon_energy}.

\subsection{Selection of photon candidates}
\label{subsec:fisher}

To combine the information contained in the two discriminating variables and define a criterion for the identification of photon candidate events, a burn sample is extracted from the set in a way that guarantees a time distribution of events representing a fair sample of the instantaneous exposure over the time period considered.
The burn sample consists of 886  events, corresponding to $\sim$1.8\% of the total selected events.
The use of such a subset avoids reliance on simulations of showers initiated by nuclei, which constitute the background for the photon search, and the related uncertainties stemming from the assumptions on the mass composition and the modeling of hadronic interactions.

\begin{figure}[t]
	\centering
    \includegraphics[width=0.49\textwidth]{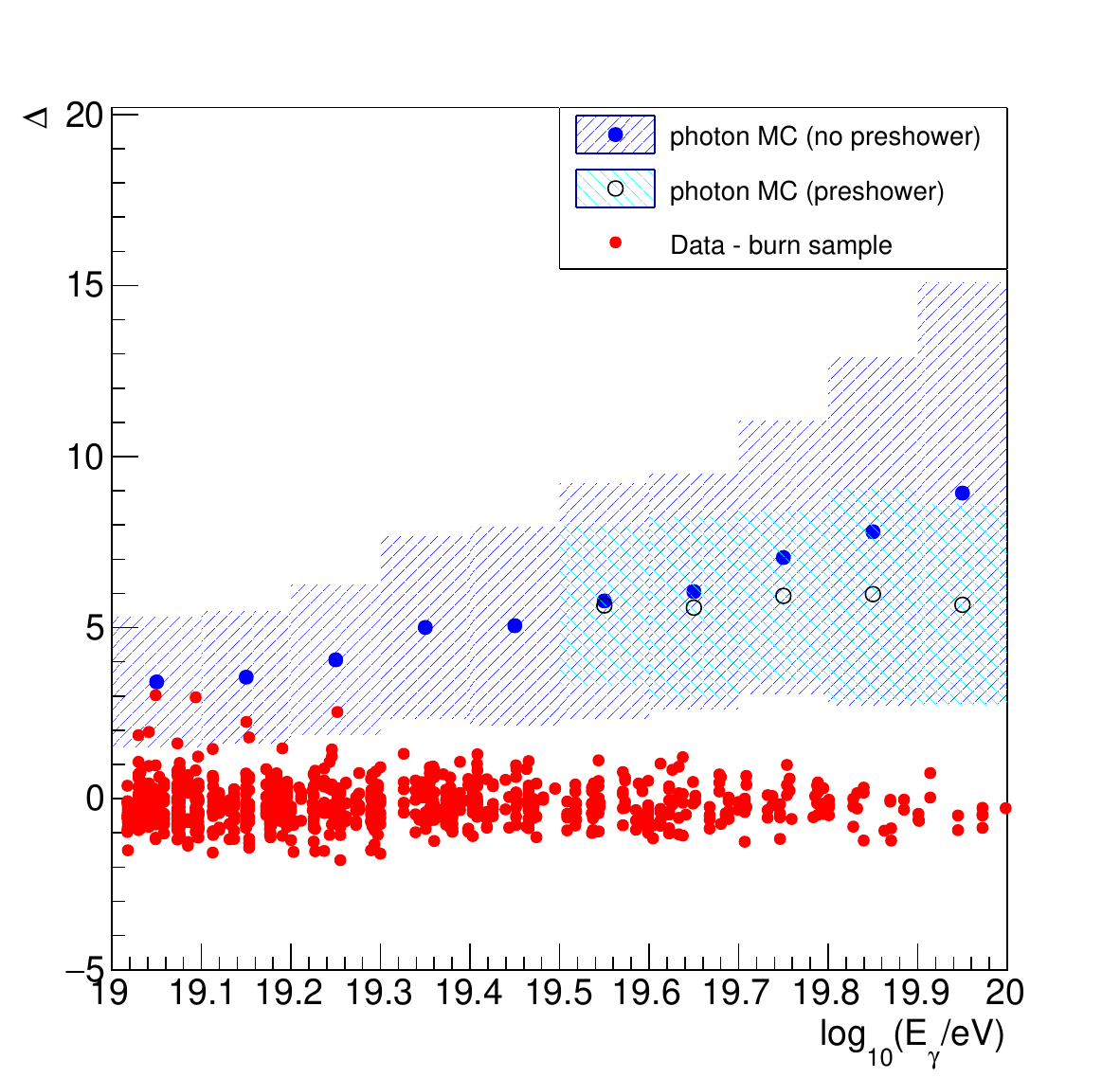}
    \includegraphics[width=0.49\textwidth]{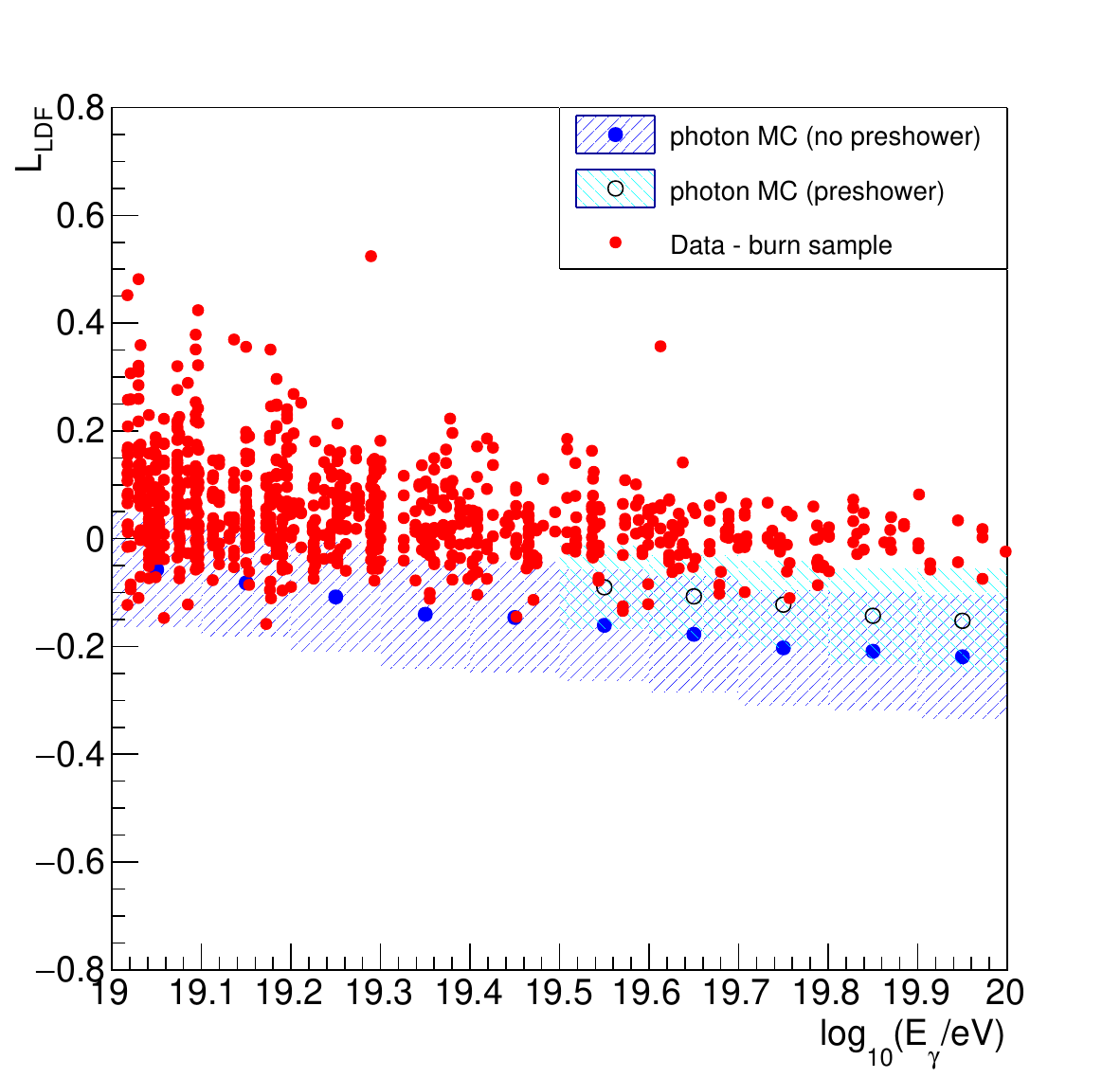}
	\caption{Distribution of $\Delta$ (left) and $L_{\rm LDF}$ (right) as a function of the photon energy $E_{\gamma}$  for the burn sample  and photon simulations. Preshowering photons are shown only in the energy range in which they represent a fraction of the selected events larger than 3\%. The bands represent one standard deviation of the photon distributions.}
	\label{fig:vardep}
\end{figure}

\begin{figure}[t]
	\centering
    \includegraphics[width=0.55\textwidth]{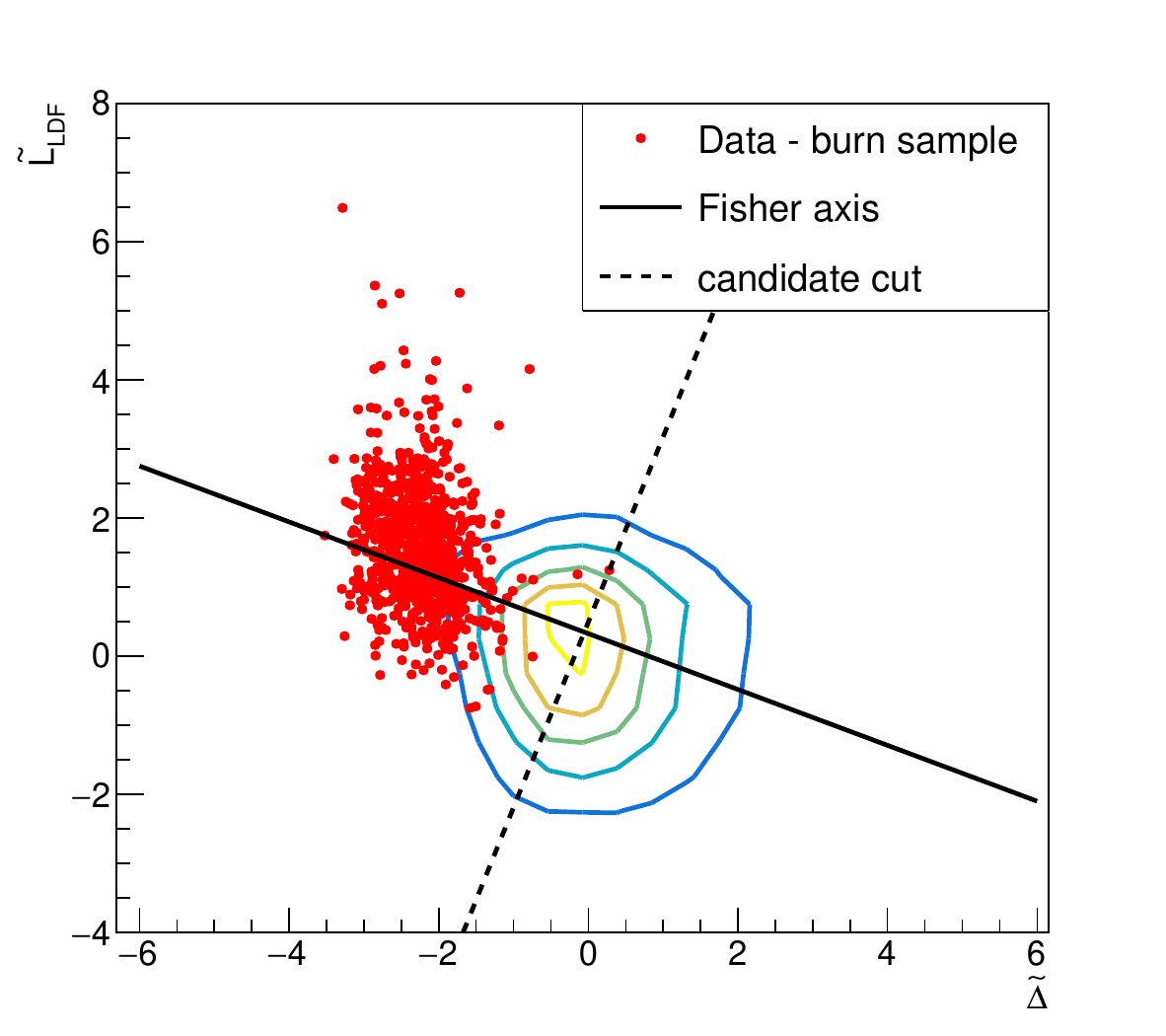}
	\caption{Distributions of the variables $\tilde{L}_{\rm LDF}$ and $\tilde{\Delta}$ of  the burn sample (points) and non-preshowering photons weighted to an $E^{-2}$ spectrum (contours). The contour levels encompass respectively 10-30-50-70-90\% of the distribution. The threshold photon energy is $10^{19}~$eV.}
	\label{fig:PCAburn2D}
    \end{figure}

For photon showers, the distribution of the SD observables $\Delta$ and $L_{\rm LDF}$
described in~\cref{sec:discrim_variables} is dependent on the energy and zenith angle of the primary particle (see, e.g., \cref{fig:vardep}). To define a single selection criterion for photon candidates as independent as possible from direction and energy,
the mean and standard deviation of the distribution of each initial variable are computed, for non-preshowering photons, in 30 bins of roughly equal statistics in the ($S(1000)$, $\theta$) space (five in $S(1000)$ and six  in $\theta$). The reference spectrum is $\propto E^{-2}$ hence simulations are weighted accordingly.
The  variables $\tilde{\Delta}$ and $\tilde{L}_{\rm LDF}$ are then defined as linear transformations of the initial ones centered around $0$ and expressed in units of the standard deviation of the corresponding distributions of non-preshowering photons:
\begin{eqnarray}
\tilde{\Delta} &=& \frac{\Delta-\langle\Delta\rangle^{i}}{\sigma_\Delta^{i}},\\
\tilde{L}_{\rm LDF} &=& \frac{L_{\rm LDF}-\langle L_{\rm LDF}\rangle^{i}}{\sigma_{L_{\rm LDF}}^{i}},
\end{eqnarray}
where $i$ is the index of the bin corresponding to the specific event. 
The distributions of $\tilde{\Delta}$ and $\tilde{L}_{\rm LDF}$ are shown in~\cref{fig:PCAburn2D}.

The two variables are combined using a Fisher discriminant analysis \cite{Fisher} with the burn sample representing the background and photon simulations the signal. The transformation is normalized  as to be equivalent to a rotation in the $(\tilde{\Delta},\tilde{L}_{\rm LDF})$ plane.
The resulting axis is shown in~\cref{fig:PCAburn2D}.

The photon candidate cut is chosen a priori as the median of the photon sample of non-preshowering events weighted to a $E^{-2}$ spectrum.
This cut value constitutes a good compromise between efficiency and purity. 
Any event falling on the right side of this cut, shown as a dashed line in~\cref{fig:PCAburn2D}, will be considered as a photon candidate.

\section{Results of the photon search}
\label{sec:results}

\begin{figure}[ht]
\centering
\includegraphics[width=0.49\textwidth]{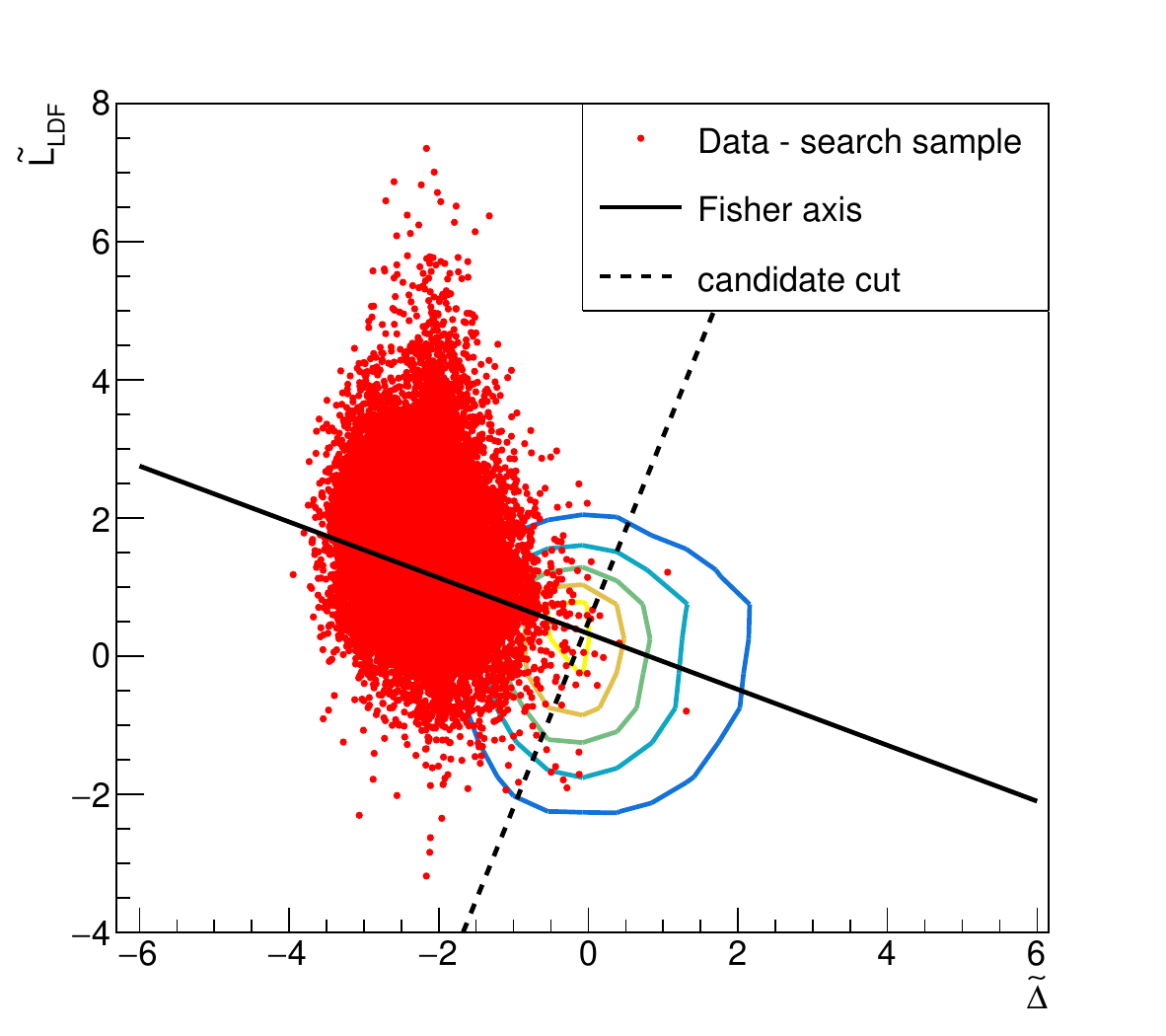}
\includegraphics[width=0.49\textwidth]{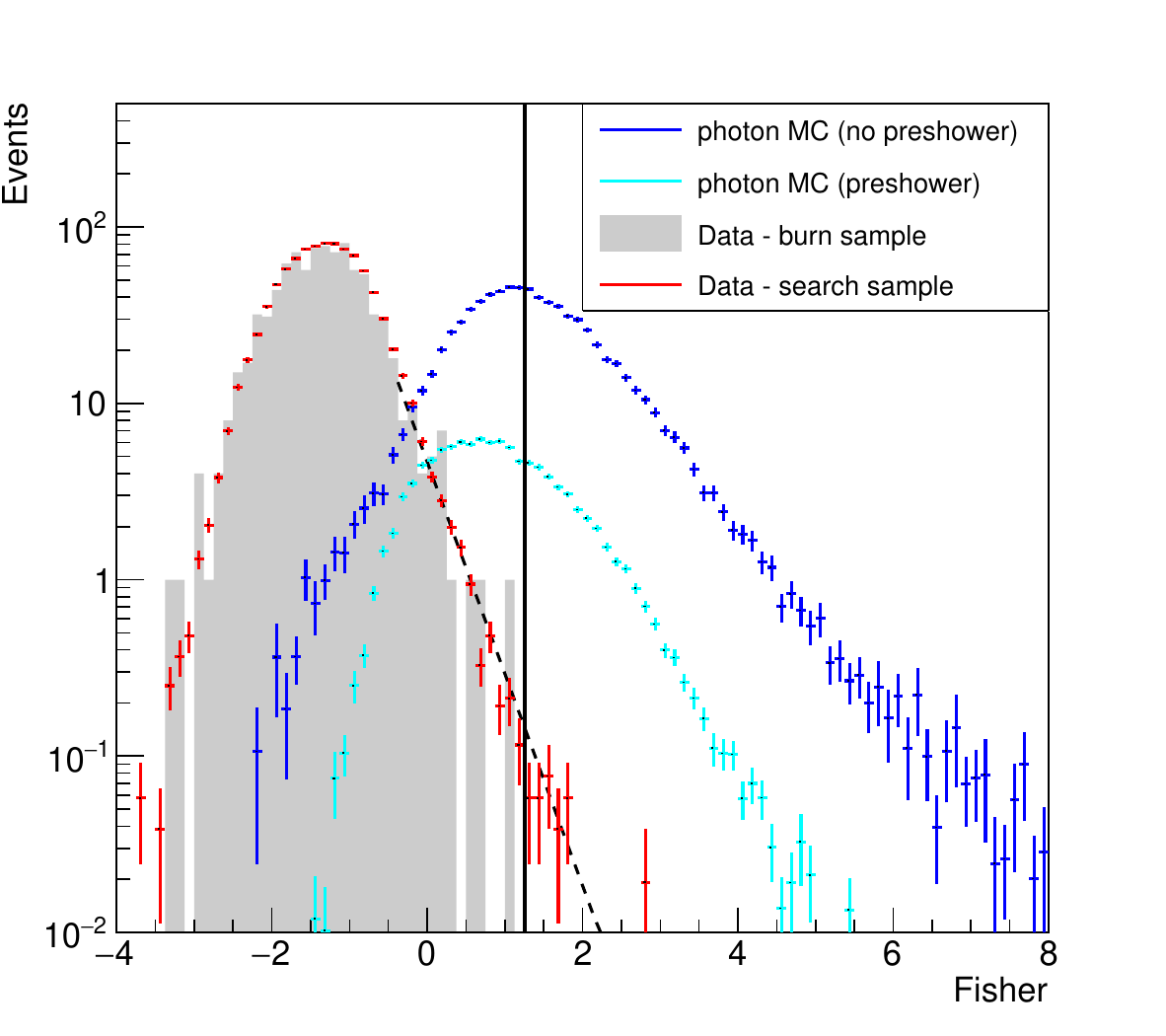}
\caption{(Left) Distribution in the $(\tilde{\Delta}$,$\tilde{L}_{\rm LDF})$ plane of the events of the search sample compared with the $E^{-2}$-weighted distribution for non-preshowering photons (same contour levels as in~\cref{fig:PCAburn2D}). (Right) Distribution of the Fisher discriminant value for the burn sample (grey), search sample (red) and $E^{-2}$-weighted photon simulations (non-preshowering in blue and preshowering in light blue). The search sample and photon distributions are scaled as to have the same integral as the burn sample one. The vertical line stands for the candidate cut. The dashed line shows the result of the exponential fit to the 5\% of events in the burn sample with larger Fisher values.}
\label{fig:PCA}
\end{figure}

Excluding the burn sample from the final analysis, the search sample consists of 48,061 selected events. Application of the photon search method yields the summary plots shown in~\cref{fig:PCA} for $E_\gamma\geq 10^{19}~$eV. Analyzing the data in the $(\tilde{\Delta}$, $\tilde{L}_{\rm LDF})$ plane results in the red points displayed in the left panel, on which are drawn the same contour levels as in~\cref{fig:PCAburn2D} of the distribution for photons as well as the Fisher axis and the candidate-cut Fisher value. In the right panel, the corresponding distributions of the Fisher discriminant value are shown as normalized histograms for the burn sample, the search sample, as well as the simulated photon sample separated in non-preshowering and preshowering. For reference, the candidate cut is shown as the vertical line, while the result of an exponential fit to the 5\% of events from the burn sample with the largest Fisher values is drawn to guide the eye in the interpretation of the tail of the Fisher distribution of the search sample. 

We find 16 (1) [0] photon candidates above $10^{19}~$eV ($2{\times}10^{19}~$eV) [$4{\times}10^{19}~$eV]. The number of observed candidates is in statistical agreement with what is expected from the exponential fit to the burn sample, with a difference of -0.3 standard deviations.
In addition, no peak-like features above the selection cut that would indicate the presence of a photon population are observed above the fall-off of the distribution. Overall, therefore, the Fisher distribution of photon candidates is consistent with the expectations of a background of UHECR events.

To search for further imprints that would be indicative of the presence of photon events, we have checked that no candidates are coincident in time. We have also searched for small-scale clustering in arrival directions that would be indicative of repeaters and thus of point-like sources of photons. No such clustering is observed, and the arrival directions of the candidates are distributed in accordance with the directional exposure of the cosmic-ray background events.

From the absence of photon flux measurements, upper limits are derived from the number of candidates above a minimum energy $E_0$. The signal efficiency of the analysis, $\epsilon_{\gamma}$, is estimated by applying the quality cuts and the selection of candidates to a Monte-Carlo set of photon air showers using the same procedure as for the data. The (weighted) ratio of the number of selected candidates (preshowering or not) reconstructed in the selected zenithal range and with assigned photon energy above the threshold $E_{0}$ to the total number of simulated events with generated zenith and energy in the desired ranges is computed. The weight is assigned so as to obtain, from the generated $E^{-1}$ spectrum, the results corresponding to the spectral shape assumed for the photon flux $\Phi(E)$ we want to bind, as explained in~\cref{subsec:Library}.
\begin{equation}\label{eqn:efficiency}
\epsilon_{\gamma}(>E_0) = \frac{ \int_{\Delta\Omega}  \int_{E_{\gamma}>E_0} \Phi_{\rm obs}(E_\gamma,\Omega_{\rm rec})\,d\Omega_{\rm rec}\,dE_{\gamma} }{ \int_{\Delta\Omega} \int_{E>E_0}\Phi_{\rm gen}(E)\,w(E)\,d\Omega\,dE}
\end{equation}
where $\Delta\Omega$ is the solid angle corresponding to the analysis selection, 
\begin{equation}
\Phi_{\rm obs}(E_\gamma,\Omega_{\rm rec}) = \int_{\Delta\Omega_{\rm gen}} \int_{\Delta E_{\rm gen}} p^{\rm cand}(E_\gamma,\Omega_{\rm rec}|E,\Omega)\,\Phi_{\rm gen}(E)\,w(E)\,d\Omega\,dE
\end{equation}
$p^{\rm cand}(E_\gamma,\Omega_{\rm rec}|E,\Omega)$ represents the probability that a shower of true Monte Carlo energy $E$ generated in the solid angle $\Omega$ is selected as candidate photon event with photon energy $E_\gamma$ and in the reconstructed solid angle $\Omega_{\rm rec}$. Here the integral is extended to the whole range of simulations in solid angle $\Delta\Omega_{\rm gen}$ and energy range $\Delta E_{\rm gen}$.
In other words, the efficiency represents the ratio between the integral spectrum folded into the detector response and the true underlying one, accounting for any effects due to inaccuracies in the photon energy and direction assignment (and possible biases for model fluxes different from $E^{-2}$).

Upper limits to the integrated flux at CL confidence level are derived using the Feldman-Cousins method~\cite{Feldman_1998}:
\begin{equation}\label{eqn:phiUL}
    \mathcal{F}^{\text{CL}}(E_\gamma\geq E_0) = \frac{N_\gamma^{\rm CL}}{A(1-\eta_{B})\epsilon_\gamma},
\end{equation}
with $A=46,900~$km$^2$~sr~yr the total exposure in the time period 1 January 2004 to 30 June 2020 and in the zenithal range considered, $\eta_{B}$ the exposure fraction corresponding to the burn sample of data used to set the candidate cut and excluded from the search,  and $N_\gamma^{\text{CL}}$ the Feldman-Cousins upper limit to the number of photon events computed at confidence level CL in the conservative hypothesis of no background event expected.

\begin{figure}[t]
\centering
\includegraphics[width=0.5\textwidth]{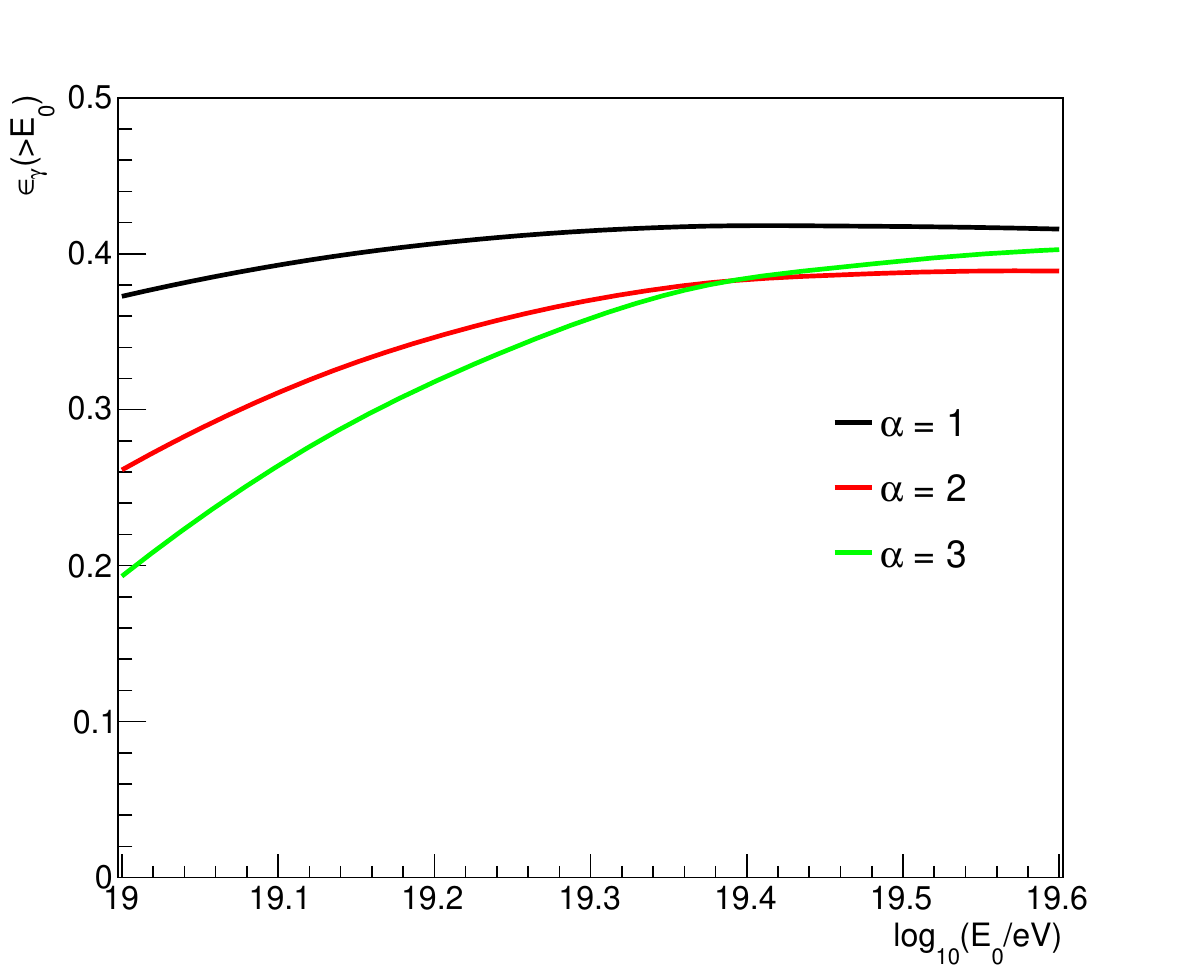}
\caption{Photon integral efficiency in the case of a power-law spectrum $E^{-\alpha}$ as function of the threshold energy $E_{0}$}
\label{fig:efficiency}
\end{figure}

While the reconstruction and candidate selection procedure is independent from the model spectrum, upper limits depend on the spectrum through the efficiency. The values obtained for both efficiencies and upper limits at 95\% CL are listed in \autoref{tab:SlopesTab} in the case of a power law spectrum $E^{-\alpha}$ for three different values of the exponent and shown in \autoref{fig:efficiency} as a function of $E_0$. The maximum value shown corresponds to $E_{\gamma} \simeq$~40~EeV where the fraction of preshowering photons in the detectable flux reaches the level of 5\% in the case of $\alpha = 2$. For larger values of $E_0$, the interpretation of the meaning of the threshold is complicated by the co-presence of two populations of photons (preshowering and not) differing in energy scale (\autoref{subsec:cuts}) and will be discussed in a future work. 
\begin{table}[ht] 
\centering
    \begin{tabular}{|c|c c c|c c c|}
      \hline
      $E_{0}$ & 10\,EeV & 20\,EeV & 40\,EeV & 10\,EeV & 20\,EeV & 40\,EeV \\
      \hline
      $\alpha$ & \multicolumn{3}{c|}{ Efficiency ($\epsilon_{\gamma}$)} & \multicolumn{3}{c|}{Flux U.L. ($10^{-3}\,{\rm km}^{-2}\,{\rm sr}^{-1}\,{\rm yr}^{-1}$)}\\
      \hline 
      1.0         & 0.37 & 0.41 & 0.42     & 1.48 & 0.273 & 0.162  \\
      2.0         & 0.26 & 0.36 & 0.39     & 2.11 & 0.312 & 0.172  \\
      3.0         & 0.19 & 0.34 & 0.41     & 2.86 & 0.332 & 0.166  \\
      \hline
    \end{tabular}
    \caption{Efficiencies and upper limits to the integral flux of photons above $E_{0}$ for different values of the exponent $\alpha$ in the case of a power-law spectrum. The limits (95\% CL) are listed on the right.}
    \label{tab:SlopesTab}
\end{table}

\begin{figure}[t]
\centering
\includegraphics[width=0.9\textwidth]{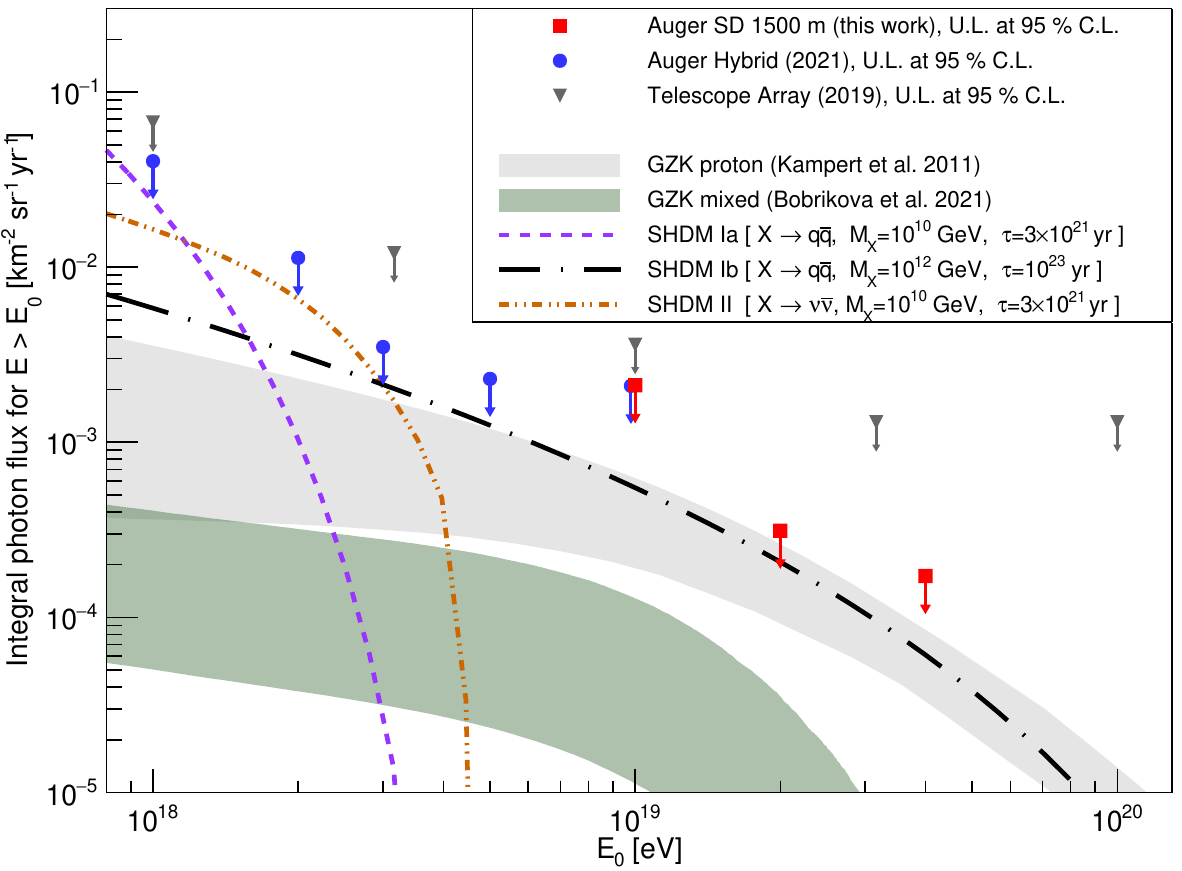}
\caption{Upper limits (at 95 \% CL) on the integral photon flux above $10^{19}$~eV determined in this study (red squares). Shown are also previous upper limits reported in~\cite{Savina:2021zrn} (Auger Hybrid, blue circles), and Telescope Array~\cite{TA_SDPhoton2019} (grey triangles). The grey band stands for the range of expected photon fluxes under the assumption of a pure-proton scenario~\cite{Sarkar_GZKphotons_ICRC11}, while the green one for the case of a mixed composition~\cite{Bobrikova:2021kuj}. In addition, several expected photon fluxes from the decay of SHDM particles are shown (lines).
The lines labeled SHDM I (case a and b) ~\cite{Kalashev_SHDM2016} and SHDM II \cite{Kachelriess_NuSHDM2018} have been obtained through private communication with one of the authors.}
\label{fig:limit}
\end{figure}

The 95\% CL upper limits to the integral photon flux with a spectral index $\alpha = 2$  are shown in~\cref{fig:limit}, together with the results of other photon searches using other components of the Observatory for lower energy
thresholds~\cite{Savina:2021zrn} and those reported from Telescope Array~\cite{TA_SDPhoton2019}. In the energy region above $10^{19}~$eV considered, the limits obtained in this study are the most stringent currently available in literature. Their astrophysical implications
are briefly discussed in \cref{sec:discussion}. 

\section{Discussion}
\label{sec:discussion}

As mentioned in~\cref{sec:intro}, the interactions of UHECRs with the background photon fields permeating the Universe, most notably the cosmic microwave background, guarantee the existence of a cosmogenic diffuse flux of UHE photons. However, the short photon horizon compared to the cosmic-ray one makes the photon flux lower than the UHECR one by several orders of magnitude. Moreover, the exact photon flux is further reduced as the mass composition of UHECRs gets heavier, because the photons produced from primary heavy nuclei are of lower energies than those from lighter ones. This is evidenced in~\cref{fig:limit} through the grey band standing for the photon flux expected from proton interactions~\cite{Sarkar_GZKphotons_ICRC11}, which is observed one order of magnitude (as well as to extend to higher energies) above that expected from a mixed composition shown in green~\cite{Bobrikova:2021kuj}. The size of the bands accounts for several hypotheses necessary for the modeling of the UHECR flux, such as the maximum acceleration energy of the nuclei at the sources and the shape of the energy spectrum of the accelerated particles. Note that the mixed-composition case is designed to reproduce the mass composition and the energy spectrum inferred from the data collected at the Observatory. The sensitivity of the present search around  $2 \times 10^{19}$~eV approaches the most optimistic expectations of the cosmogenic photon flux from protons, while it is one order of magnitude above those from the mixed-composition model. In the future, it will thus be possible to probe the proton expectations with improved sensitivity to photons, and thus to confirm or reject indirectly the presence of a mixed composition at the highest energies.

Our upper limits have also consequences for models of super-heavy dark matter (SHDM).
Although the absence of a detectable flux of UHE photons at a level comparable to that of cosmic rays excludes that SHDM particles could explain the bulk of UHECRs, they can still contribute in a subdominant way to the all-particle spectrum. For a $E^{-2}$ photon spectrum, and using the flux of cosmic rays measured using  the Pierre Auger Observatory \cite{Aab:2020gxe}, the upper limits (at 95\% CL) to the integral photon fraction are: 1.6\%, 1.2\% and 3.2\% above 10~EeV, 20~EeV and 40~EeV respectively.

There are several theoretical motives to search for SHDM related to particle physics in the early Universe. A recent exploration of the constraints provided by our photon-flux limits on some  models is given elsewhere~\cite{PierreAugerCollaboration:2022tlw}. Here, we restrict ourselves to illustrate in~\cref{fig:limit} the discovery potential with searches for UHE photons: we show as the dashed violet line and the dot-dashed red line the expected photon fluxes in the case of hadronic~\cite{Kalashev_SHDM2016} and leptonic~\cite{Kachelriess_NuSHDM2018} decay channels, for a mass of the SHDM particles of $10^{10}~$GeV and a lifetime of $3{\times}10^{21}~$yr. In addition, we also show as the black dot-dashed line an allowed flux extending to higher energy for a mass of $10^{12}~$GeV and a lifetime of $10^{23}~$yr (hadronic decay channel~\cite{Kalashev_SHDM2016}). As the sensitivity of current photon searches increases, it will be possible to further constrain the allowed values of mass and lifetime.

In summary, a search for photons using the surface detector of the Pierre Auger Observatory has been performed, leading to the most stringent limits on the diffuse photon flux above $10^{19}$~eV and complementing the limits obtained in the lower energy decade.  The current limits are not able to challenge the 
flux of cosmogenic photons expected for the mixed-composition model but start to probe the most optimistic predictions of the pure-proton scenarios.
Future data will enable us to constrain further the different mechanisms expected to produce UHE photons. 
In particular, the completion of the upgraded Observatory~\cite{Aab:2016vlz} will allow us to increase the sensitivity of the different discriminating analyses.  
Finally, the analysis presented in this study can also serve for directional searches from specific targets or searches in time coincidence with observations from other messengers.

\section*{Acknowledgments}

\begin{sloppypar}
The successful installation, commissioning, and operation of the Pierre
Auger Observatory would not have been possible without the strong
commitment and effort from the technical and administrative staff in
Malarg\"ue. We are very grateful to the following agencies and
organizations for financial support:
\end{sloppypar}

\begin{sloppypar}
Argentina -- Comisi\'on Nacional de Energ\'\i{}a At\'omica; Agencia Nacional de
Promoci\'on Cient\'\i{}fica y Tecnol\'ogica (ANPCyT); Consejo Nacional de
Investigaciones Cient\'\i{}ficas y T\'ecnicas (CONICET); Gobierno de la
Provincia de Mendoza; Municipalidad de Malarg\"ue; NDM Holdings and Valle
Las Le\~nas; in gratitude for their continuing cooperation over land
access; Australia -- the Australian Research Council; Belgium -- Fonds
de la Recherche Scientifique (FNRS); Research Foundation Flanders (FWO);
Brazil -- Conselho Nacional de Desenvolvimento Cient\'\i{}fico e Tecnol\'ogico
(CNPq); Financiadora de Estudos e Projetos (FINEP); Funda\c{c}\~ao de Amparo \`a
Pesquisa do Estado de Rio de Janeiro (FAPERJ); S\~ao Paulo Research
Foundation (FAPESP) Grants No.~2019/10151-2, No.~2010/07359-6 and
No.~1999/05404-3; Minist\'erio da Ci\^encia, Tecnologia, Inova\c{c}\~oes e
Comunica\c{c}\~oes (MCTIC); Czech Republic -- Grant No.~MSMT CR LTT18004,
LM2015038, LM2018102, CZ.02.1.01/0.0/0.0/16{\textunderscore}013/0001402,
CZ.02.1.01/0.0/0.0/18{\textunderscore}046/0016010 and
CZ.02.1.01/0.0/0.0/17{\textunderscore}049/0008422; France -- Centre de Calcul
IN2P3/CNRS; Centre National de la Recherche Scientifique (CNRS); Conseil
R\'egional Ile-de-France; D\'epartement Physique Nucl\'eaire et Corpusculaire
(PNC-IN2P3/CNRS); D\'epartement Sciences de l'Univers (SDU-INSU/CNRS);
Institut Lagrange de Paris (ILP) Grant No.~LABEX ANR-10-LABX-63 within
the Investissements d'Avenir Programme Grant No.~ANR-11-IDEX-0004-02;
Germany -- Bundesministerium f\"ur Bildung und Forschung (BMBF); Deutsche
Forschungsgemeinschaft (DFG); Finanzministerium Baden-W\"urttemberg;
Helmholtz Alliance for Astroparticle Physics (HAP);
Helmholtz-Gemeinschaft Deutscher Forschungszentren (HGF); Ministerium
f\"ur Kultur und Wissenschaft des Landes Nordrhein-Westfalen; Ministerium
f\"ur Wissenschaft, Forschung und Kunst des Landes Baden-W\"urttemberg;
Italy -- Istituto Nazionale di Fisica Nucleare (INFN); Istituto
Nazionale di Astrofisica (INAF); Ministero dell'Istruzione,
dell'Universit\'a e della Ricerca (MIUR); CETEMPS Center of Excellence;
Ministero degli Affari Esteri (MAE); M\'exico -- Consejo Nacional de
Ciencia y Tecnolog\'\i{}a (CONACYT) No.~167733; Universidad Nacional Aut\'onoma
de M\'exico (UNAM); PAPIIT DGAPA-UNAM; The Netherlands -- Ministry of
Education, Culture and Science; Netherlands Organisation for Scientific
Research (NWO); Dutch national e-infrastructure with the support of SURF
Cooperative; Poland -- Ministry of Education and Science, grant
No.~DIR/WK/2018/11; National Science Centre, Grants
No.~2016/22/M/ST9/00198, 2016/23/B/ST9/01635, and 2020/39/B/ST9/01398;
Portugal -- Portuguese national funds and FEDER funds within Programa
Operacional Factores de Competitividade through Funda\c{c}\~ao para a Ci\^encia
e a Tecnologia (COMPETE); Romania -- Ministry of Research, Innovation
and Digitization, CNCS/CCCDI UEFISCDI, grant no. PN19150201/16N/2019 and
PN1906010 within the National Nucleus Program, and projects number
TE128, PN-III-P1-1.1-TE-2021-0924/TE57/2022 and PED289, within PNCDI
III; Slovenia -- Slovenian Research Agency, grants P1-0031, P1-0385,
I0-0033, N1-0111; Spain -- Ministerio de Econom\'\i{}a, Industria y
Competitividad (FPA2017-85114-P and PID2019-104676GB-C32), Xunta de
Galicia (ED431C 2017/07), Junta de Andaluc\'\i{}a (SOMM17/6104/UGR,
P18-FR-4314) Feder Funds, RENATA Red Nacional Tem\'atica de
Astropart\'\i{}culas (FPA2015-68783-REDT) and Mar\'\i{}a de Maeztu Unit of
Excellence (MDM-2016-0692); USA -- Department of Energy, Contracts
No.~DE-AC02-07CH11359, No.~DE-FR02-04ER41300, No.~DE-FG02-99ER41107 and
No.~DE-SC0011689; National Science Foundation, Grant No.~0450696; The
Grainger Foundation; Marie Curie-IRSES/EPLANET; European Particle
Physics Latin American Network; and UNESCO.
\end{sloppypar}

\bibliography{main}
\bibliographystyle{JHEP}

\end{document}